\documentstyle[12pt,epsfig]{article}
\newlength{\dinwidth}
\newlength{\dinmargin}
\newcommand{\ba}{\begin{array}}
\newcommand{\ea}{\end{array}}
\newcommand{\be}{\begin{equation}}
\newcommand{\ee}{\end{equation}}
\newcommand{\bea}{\begin{eqnarray}}
\newcommand{\eea}{\end{eqnarray}}
\newcommand{\gsim}{\mathrel{\mathop{\kern 0pt \rlap
  {\raise.2ex\hbox{$>$}}} \lower.9ex\hbox{\kern-.190em $\sim$}}}

\def\log{{\rm{log}}}

\def\ben{\begin{equation}}
\def\een{\end{equation}}
\def\bea{\begin{eqnarray}}
\def\eea{\end{eqnarray}}

\begin{document}
\thispagestyle{empty}
\addtocounter{page}{-1}
\vskip-0.35cm
\begin{flushright}
IFT 04/10\\
{\tt hep-th/0403250}
\end{flushright}
\vspace*{0.2cm}
\centerline{\Large \bf IR Renormalons and Fractional Instantons}
\vskip0.3cm
\centerline{\Large \bf in SUSY Gauge Theories}

\vspace*{1.0cm} 
\centerline{\bf Cesar Gomez}
\vspace*{0.7cm}

\centerline{\it Instituto de Fisica Teorica CSIC/UAM, C-XVI Universidad Autonoma,}
\vspace*{0.2cm}
\centerline{\it E-28049 Madrid \rm SPAIN}
\vspace*{1cm}
\centerline{\tt Cesar.Gomez@uam.es}

\vspace*{0.8cm}
\centerline{\bf abstract}
\vspace*{0.3cm}
We study IR-renormalon divergences in $N=1$ supersymmetric Yang Mills gauge theories and in
two dimensional non linear sigma models with mass gap. We derive, in both types of theories, a direct
connection between IR- renormalons and fractional instanton effects. From the point of view of large $N$
dualities we work out a connection between IR-renormalons and $c=1$ matrix models.
\baselineskip=18pt
\newpage
\section{Introduction}
Quantum Field Theory is defined in terms of perturbative expansions
in the coupling constant. Even for asymptotically free theories and
independently how small is the coupling constant these perturbative expansions
are, for theories with no massive fields, divergent. To study the nature
and meaning of these divergences is the most natural perturbative window
into non perturbative physics.

In the large $N$ limit \cite{thooft1} the divergences of the perturbative 
planar expansion are less severe due to the fact that the number
of Feynman diagrams at n- loop order only grow geometrically \cite{koplic}. This in
particular means that instanton divergences are absent in the large $N$ planar limit. However
in the planar limit other type of divergences survive. These are known as renormalons
\cite{parisi}, \cite{thooft2}, \cite{thooft3} and can be associated with n-loop Feynman diagrams
that behave thenselves as $n!$. Mathematically IR-renormalons manifest as singularities
of the Borel transform of Green functions. These singularities prevent Borel summability
in a direct way. Some time ago 't Hooft suggested that IR- renormalons could be important
in the confinement dynamics and in the generation of a mass gap.

On the other hand the study of the dynamics of $N=1$ super Yang Mills  seems to
point out to some sort of {\em ``fractional instanton''} as the the dynamical origin of
mass generation as well as of non vanishing chiral condensates. A fractional
instanton is a formal concept and unfortunately we dont have, in the infinite volume limit, classical
gauge configurations with fractional topological number \footnote{ In the finite volume we
have toron solution \cite{thooft4}}.
However it is clear why fractional instantons could be relevant, namely and contrary
to the case of instantons they survive in the large $N$ planar limit.

In this paper we will provide considerable evidence that for $N=1$ gauge theories
{\em fractional instantons are in fact IR renormalons}. The way we will proceed is
by evaluating the Borel transform in a stationary phase approximation. As a check we
will work out some two dimensional non linear sigma models where the existence of a mass gap
can be proved in the large $N$ approximation. 

An intimately related question is the already old discussion on the $\theta$ dependence
of physics in the large $N$ limit \cite{wittentheta}, \cite{veneziano}. The mass of
the $\eta'$ as well as the topological susceptibility for pure Yang Mills are
examples of effects that in finite $N$ we could associate with instantons but that
we expect survive in the large $N$ limit at order $\frac{1}{N}$. The understanding of these effects
in the large $N$ is very much based on our experience with two dimensional sigma
models with dynamical generation of a mass gap \cite{Dadda}, \cite{wittemcp}. As already mentioned we will
work out some IR- renormalon effects for these models and we will discuss their relevance
in the mass gap generation.

A different approach to the physical meaning of
renormalon divergences for the planar perturbative expansion
is in the context of large N dualities \cite{thooft1}. For a gauge theory based on a gauge group
of rank $N$ the pertubative expansion is organized as a genus string expansion: 
$\sum_{l} g_{s}^{2l-2} F_{l}(t)$ where $l$ is the genus, $t=g^{2}N$ is 't Hooft coupling
and where the ``string'' coupling $g_{s}$ is $g^{2}$. The dual string hypothesis is that in the
large $N$ the perturbative loop expansion in 't Hooft coupling $t$ can be summed and interpreted
as the closed string amplitude in some target space time characterized by $t$. For asymptotically free
theories we should expect divergences of $F_{0}(t)$, governed by IR renormalons, 
independently how small is $t$. In some recent papers \cite{vafaooguri} and for simple traget space
geometries related with the conifold, it was suggested that singularities at $t=0$ could be related
with fractional instanton effects. 
We observe that for asymptotically free theories the uncertainties due to 
renormalon ( and instanton) divergences in the perturbative expansion can be interpreted as 
associated with
a hidden gravitational sector in the sense of Matrix $c=1$ models. This can be related with
t' Hooft old suggestion that asymptotic freedom is not enough due 
to the uncertainties in the perturbative expansion even for arbitrarily small coupling.
Probably the new ingredient we need in order to fix these uncertainties is intimately related with
a hidden gravitational sector in asymptotically free gauge theories.

The plan of the paper is as follows. In the first section we mainly review some aspects of
the use of instantons in supersymmetric QCD and we point out on the difficulties for extending
this type of approach in the large $N$. In the second section we present our approach
to IR renormalons and their connection with fractional instantons. In the next section we work
out some two dimensional sigma models with mass gap. Finally we make some comments on the stringy
interpretation in the context of large $N$ dualities ending with a short remark on the connection
of renormalons and Matrix models. 

\section{Instantons and $N=1$ super Yang Mills}
Asymptotically free theories with no infrared fixed point are
expected to generate a mass gap dynamically. In the presence of
massless fermions this is expected to be accompanied by chiral
symmetry breaking. Unfortunately, the analytic tools available
for studying these phenomena are rather limited. In the 80's and
90's considerable progress has been made, however, in supersymmetric gauge 
theories - thanks to holomorphy and nonrenormalization properties.

One important example is $N=1$ chromodynamics \cite{affleck}, \cite{vsz1} with gauge group
$SU(N_c)$ and with $N_f$ flavors and $N_c \geq N_f + 1$. 
From general arguments it follows that
nonperturbatively the theory generates a superpotential. This is a
$F$ term whose form is
\ben
F = b g^4 \Lambda_{SQCD}^{3N_c - N_f \over N_c - N_f}
\int d^2\theta ({\rm det}_{rr'} {\bar Q}_{ir} Q^{ir'})^{-{1\over N_c-N_f}}
\label{eq:one}
\een
where $\Lambda_{SQCD}$ is the scale generated by the beta function. At one loop
this is given in terms of a cutoff $\Lambda_0$ and the bare coupling
$g$ by
\ben
\Lambda_{SQCD} = \Lambda_0 e^{-{8\pi^2 \over (3N_c -N_f)g^2}}
\label{eq:two}
\een
At the classical level the theory has moduli given in terms of the expectation
values of the squarks. We will restrict our attention to 
the case where the expectation values are
symmetric in flavor indices and denoted by $v$.
Then F-term implies a mass for the fermions
\ben
m \sim   \Lambda_{SQCD}^{3N_c - N_f \over N_c - N_f}~v^{-{2N_c \over
N_c-N_f}} \sim \Lambda_{0}^{3N_c - N_f \over N_c - N_f}~v^{-{2N_c \over
N_c-N_f}}~e^{-{8\pi^2 \over (N_c-N_f)g^2}}
\label{eq:three}
\een
Because of the F-term, the flat direction is however broken and in
fact the true minimum is at infinity. 

To make sense of the theory it is customary to add an explicit mass
term with mass parameter $M$ to the action. Then the total potential has
a local minimum at
\ben
v = [{g^4 N_f \over M(N_c-N_f)}]^{N_c-N_f \over 2N_c}~\Lambda^{3N_c-N_f 
\over 2N_c}
\label{eq:four}
\een
This means that one has a well defined calculation for the gluino
condensate $<\lambda \lambda>$ by using the Konishi anomaly
\ben
<\lambda \lambda > = {1\over g^2} M~v^2
\een
which leads to, using (\ref{eq:four})
\bea
<\lambda \lambda> & \sim & {g^2N_f \over N_c-N_f}
\Lambda^{3N_c - N_f \over N_c - N_f}~
v^{-{2N_f \over N_c - N_f}} \nonumber \\
	& \sim & ({N_f \over N_c-N_f})^{N_c-N_f \over N_c}~M^{N_f \over N_c}~
\Lambda^{3N_c - N_f \over N_c}
\label{eq:five}
\eea
In fact these results may be used to learn about supersymmetric
Yang-Mills theory {\em without} any matter. This is obtained as a limit
of super-QCD by taking a limit $M >> \Lambda_{SQCD}$. One can then
take $M$ to be the cutoff scale at which the bare couplings ared defined.
At energies much smaller than the cutoff the matter decouples leaving 
a pure super Yang-Mills theory. Using the one loop beta functions
of the two theories one readily derives the following expression for
the dynamically generated scale of the super-Yang-Mills, $\Lambda_{SYM}$
is given by
\ben
\Lambda_{SYM} = (M^{N_f} \Lambda_{SQCD}^{3N_c-N_f})^{1\over 3N_c}
\label{eq:six}
\een
The expression (\ref{eq:five}) then becomes
\ben
<\lambda\lambda> = \Lambda_{SYM}^3
\label{eq:seven}
\een
which is now an expression in the pure Yang-Mills theory.

While the results above are quite general and valid for any $N_f, N_c$
with $N_c \geq N_f - 1$, and follows from general symmetry properties 
together with holomorphy, not much is known about whether some
specific configurations of gauge fields are primarily responsible for
these expressions, except when $N_c = N_f -1$.
For $N_f = N_c -1$ instantons are, in fact
responsible for mass gap, chiral symmetry breaking and
gluino condensation. 

The instanton calculations are performed in the
absence of the explicit mass term and expressed in terms of the
running coupling constant $g^2(v)$ defined at the scale set by the
modulus. This is related to $\Lambda_{SQCD}$ by
\ben
\Lambda_{SQCD} = v~e^{-{8\pi^2 \over (3N_c - N_f)g^2(v)}}
\label{eq:eight}
\een
which is valid for any $N_c,N_f$ with $N_c > N_f$.
The expression for the fermion mass following from the superpotential
may be the written as
\ben
m \sim v ~e^{-{8\pi^2 \over g^2 (N_c-N_f)}}
\label{eq:ten}
\een
A standard dilute gas computation using constrained instantons would
however lead to
\ben
m \sim v ~e^{-{8\pi^2 \over g^2}}
\label{eq:tena}
\een
which would agree with (\ref{eq:ten}) if $N_c = N_f + 1$. Indeed, as is
well known, for this case instanton methods are reliable, provided one
always has $v >> \Lambda_{SQCD}$.

We will be interested in the large $N_c$ limit where
$g \rightarrow 0, N \rightarrow \infty$ with $g^2 N_c = {\rm fixed}$. 
In this limit, instantons have a large $O(N)$ action and do not contribute.
It may appear puzzling, however, that the final results for $<\lambda\lambda>$ 
for the supersymmetric Yang-Mills theory (equation (\ref{eq:six}),
(\ref{eq:seven}) remain finite in this limit. However a closer examination
quickly reveals that the relations above imply that $v \sim \Lambda_{SQCD}$
at large N, which invalidates the instanton calculation.

What, then, is the mechanism for dynamical mass generation and gluino
condensation in the large-N limit ? A similar question can be asked in
other asymptotically free theories where instanton physics is known to
account for dynamical behavior at finite $N$. 

Many years ago, 't Hooft showed that asymptotically free field theories
have {\em infrared renormalon} singularities in the Borel transform of
Green's function and suggested that these singularities play an important
role in mass generation. Typically the Borel transform for these theories 
has a rich singularity structure. Instantons, for example, manifest
themselves as singularities on the positive real axis. IR renormalons -
whose location are determined by the beta function - are distinct from
these. Indeed, in the large-N limit the instanton singularities are absent 
while the IR renormalons survive.

\section{Fractional Instantons and IR renormalons}
Let us start with a brief review of t Hoof's standard approach to instantons
\cite{thooft5}. For simplicity
we will first consider the case of pure Yang Mills without fermions. The partition
function is
\ben\label{cero2}
Z(g^{2}) = \int dA e^{-\frac{1}{g^{2}} \int dx^{4} L(A)}
\een
where we assume that gauge fixing and ghost terms has been included. Up to numerical factors
what we get after taking into account the zero modes is ( for $SU(2)$ gauge theory )
\ben\label{uno2}
Z(g^{2}) = \int dx^{4} \frac{d\rho}{\rho^{5}} \frac{1}{g^{8}} e^{- \frac{8 \pi^{2}}{g^{2}(\frac{1}{\rho})}}
\een
where now the integration over $x$ correspond to the moduli of translations and the integration
over $\rho$ to the moduli of dilatations. The factor $\frac{1}{g^{8}}$ comes from the $8$ zero modes.
From (\ref{uno2}) we get the effective action
\ben\label{dos2}
L_{eff}(x;g^{2}) dx^{4} = dx^{4} \int \frac{d\rho}{\rho^{5}} \frac{1}{g^{8}} 
e^{- \frac{8 \pi^{2}}{g^{2}(\frac{1}{\rho})}}
\een
In general for $SU(N)$ we get in (\ref{dos2}) a factor
\ben
\frac{1}{g^{4Nk}}
\een
for $k=1$ the instanton number.
Notice that the effective lagrangian (\ref{dos2}) is a dimension $4$ operator as it should be.

Let us now formally consider ``fractional instantons'' of topological number $k= \frac{1}{N}$.
If we extend to fractional topological charge the index theorem
we get as the number of zero modes $4 \frac{1}{N} N$ i.e just the four translations, independently of the
rank of the gauge group. Notice that for topological charge $\frac{1}{N}$
we have not dilatation zero modes. Extending to this case the instanton result we get
\ben\label{tres}
dx^{4} \frac{1}{g^{4}} e^{- \frac{8\pi^{2}}{g^{2}(\mu)N}} \mu^{\frac{11}{3}}
\een
Contrary to the instanton case we don't get naturally a dimension four operator and
therefore
we should include some extra scale factor in order to get an operator 
that can qualifies as an effective action \footnote{It is amussing to notice that the problem
with dimensions is in a certain sense very small and $12$ instaed of $11$ will do the job.}. 
Notice also that the
problem appears because the fractional instanton has not dilatation zero modes. All these problems in
addition to the crucial fact that we have not any concrete classical solution with fractional
topological number makes quite problematic the interpretation of fractional instanton effects in
a semiclassical approximation.

The situation becomes a bit better if we have $N=1$ supersymmetry. In this case we
should take into account the fermionic zero modes for the gluino. We can
expect an effective action
\ben\label{tres2}
L_{eff}(g^{2}) dx^{4}d\theta^{2} =
dx^{4}d\theta^{2} N  \frac{1}{g^{4}}\Lambda^{3} e^{- \frac{8 \pi^{2}}{g^{2}(\frac{1}{\Lambda})N}}
\een
that now as it is an operator of dimension $3$ qualifies as an $F$ term effective supersymmetric lagrangian.
This is in fact the reason fractional instantons are normaly invoqued in the dynamics of $N=1$ super Yang Mills.
It is interesting to notice, already from 
the previous discussion, that fractional instantons are naturally related with supersymmetry. 
This is not very surprising since we know that due to supersymmetric Ward identities the
gluino condensate associated with instanton effects can be -if formal cluster arguments are invoqued-
factorized into fractional instanton contributions.

\subsection{Renormalons}

\subsubsection{Borel Transform and Classical Configurations}
As pointed out by t Hooft we can formally relate 
the partition function (\ref{cero2}) with a Borel transform. In fact let us formally define a new variable
$z$ by $S(A) =z$ and let us denote $A(z)$ the corresponding gauge configuration. We get
\ben
Z(g^{2}) = \int_{M} \int_{0}^{\infty} dz (\frac{\partial S}{\partial z})_{z}^{-1} e^{- \frac{z}{g^{2}}}
\een
where $\int_{M}$ represents the integral over the moduli of unequivalent gauge configurations $A(z)$
with $S(A)=z$. Comparing with a Borel transform
\ben
Z(g^{2}) = \int_{0}^{\infty} dz F(z) e^{- \frac{z}{g^{2}}}
\een
it is clear that the Borel transform $F(z)$ is singular at the classical instanton solution for
$z= 8\pi^{2}$. In general we will get singularities in the Borel plane for any classical configuration. 
Notice
that we are working with euclidean signature and that the only relevant singularities for
the Borel transform are classical euclidean configurations with $S(A)=z$ positive.

\subsubsection{IR-Renormalons}
As it is well known
renormalons are divergences of perturbative expansions where the $n$ loop contribution grows as $n!$.
This $n!$ comes from the contribution of a $n$ loop diagram itself and not as it is the case
for instanton divergences from the number of diagrams contributing at the $n$ loop order. This is the
reason renormalons survive in the large $N$ limit
where it is known that the number of diagrams at order $n$ grows at most geometrically. Let us start 
considering the
following formal perturbative expansion,
\ben\label{cuatro2}
G(g^{2}) = \sum_{n=0}^{n = \infty} a_{n} g^{2n}
\een
where
\ben
a_{n} = a^{n} n!
\een
for some coefficient $a$ that we will discuss in a moment.
The Borel transform of (\ref{cuatro2}) is
\ben\label{cinco2}
G(g^{2}) = \frac{1}{g^{2}} \int_{0}^{\infty} dz \sum_{n=0}^{n=\infty} \frac{a_{n}}{n!} z^{n} e^{-\frac{z}{g^{2}}}
\een
Using (\ref{cinco2}) we observe that the renormalon divergence is at $z = \frac{1}{a}$. Notice
that we are considering the existence of a g-independent piece in $G(g^{2})$ given by $a_{0}$.

Now we will proceed to approximate (\ref{cinco2}) using stationary phase approximation.
In order to do that we rewrite (\ref{cinco2}) as
\ben
G(g^{2}) = \int_{0}^{\infty} dz e^{f(z) - \frac{z}{g^{2}}}
\een
The stationary phase approximation is given by
\ben
G(g^{2}) = e^{f(z(g^{2})) - \frac{z(g^{2})}{g^{2}}}
\een
where $z(g^{2})$ is the solution to $f'(z) = \frac{1}{g^{2}}$. Using that
\ben
\sum_{n=0}^{n=\infty} a^{n} z^{n} = e^{f(z)}
\een
we get 
\ben\label{seis2}
z(g^{2})= \frac{1}{a} -g^{2}
\een
and therefore
\ben
G(g^{2}) \sim \frac{1}{g^{4}} \frac{1}{a} e^{-\frac{1}{g^{2}a}}
\een
Notice from (\ref{seis2}) that the renormalon singularity $z= \frac{1}{a}$ correspond to $g^{2} = 0$. This
is already telling us that renormalon divergences are important at weak coupling.
Let us now discuss the physical meaning of renormalons. As we have already mention 
the renormalon singularity comes from the contribution of diagrams that
behave as $n!$. 
In order to define the renormalon we will start considering the insertion of chains of vacuum bubbles
into a propagator $P(k^{2})$. Denoting the result $P_{R}(k^{2})$ we get
\ben
P_{R}(k^{2}, \Lambda) = \sum_{n} \frac{1}{k^{2}} (ln(\frac{\Lambda^{2}}{k^{2}}))^{n} C^{n}
\een
where $C= -\frac{\beta_{1}}{2} g^{2}$ and $\beta_{1}$ is the one loop beta function. We will be interested
in IR renormalons i.e low momentum $k< \Lambda$ for $\Lambda$ a cutoff.

The IR renormalon will be defined as
\ben\label{siete2}
G_{\alpha}(g^{2}) = \int_{0}^{\Lambda} d^{4}k P_{R}(k^{2}) k^{2(1- \alpha)} =
\sum_{n}\int_{0}^{\Lambda} d^{4}k \frac{1}{k^{2\alpha}}(ln(\frac{\Lambda^{2}}{k^{2}}) )^{n}C^{n}
\een
Changing variables $ln(\frac{1}{k'^{2}}) = x$ with $k' = k \Lambda$ we get
\ben
G_{\alpha}(g^{2}) = \sum_{n}\Lambda^{4-2\alpha} \frac{1}{(2- \alpha)} \frac{C^{n}}{(\alpha - 2)^{n}} n!
\een
with $\alpha < 2$. Denoting
\ben\label{ocho2}
a= \frac{-\beta_{1}}{2(\alpha-2)} = \frac{\beta_{1}}{d}
\een
with $d= 4-2\alpha$ we get
\ben
G_{\alpha} (g^{2}) = \frac{\Lambda^{4-2\alpha}}{(2-\alpha)} a^{n} n!g^{2n}
\een
In terms of the Borel transform we get
\ben\label{nueve2}
G_{\alpha}(g^{2}) = \frac{\Lambda^{4-2\alpha}}{(2-\alpha)} \frac{1}{g^{2}} \int_{0}^{\infty} dz
F(z) e^{-\frac{z}{g^{2}}}
\een
for $F(z) = \sum_{n} a^{n}z^{n}$. 

Notice that 
\ben
G_{\alpha} (g^{2}) = \frac{\Lambda^{4-2\alpha}}{(2-\alpha)} + 
\frac{\Lambda^{4-2\alpha}}{(2-\alpha)} a g^{2} + ....
\een
If we substract the $g$- independent piece $\frac{\Lambda^{4-2\alpha}}{(2-\alpha)}$ we obtain
\ben
G_{\alpha}^{S}(g^{2}) = \frac{\Lambda^{4-2\alpha}}{(2-\alpha)} \int_{0}^{\infty} dz
F(z) e^{-\frac{z}{g^{2}}}
\een
for $G_{\alpha}^{S}(g^{2})$ the substracted Green function.

Let us now consider the meaning of $G_{\alpha}(g^{2})$. We can associate
$P_{R}(k^{2})$ with 
\ben
<\Phi(k)\Phi(-k)>
\een
where $\Phi$ represents a local quantum field of dimension one and where
we are summing all vacuum bubble chains insertions. In this sense
$G_{\alpha=1}(g^{2}) = \int_{0}^{\Lambda} d^{4}k<\Phi(k)\Phi(-k)>$ and
generically for $\alpha = 1- n$ with $\alpha <2$
\ben
G_{\alpha} (g^{2}) = \int_{0}^{\Lambda} d^{4}k<\partial_{1}..\partial_{n}\Phi(k)\partial_{1}
...\partial_{n}\Phi(-k)>
\een
where it is implicit the condition of Lorentz scalar and gauge singlet.

Using (\ref{ocho2}) and (\ref{nueve2}) we get for $\alpha = 1-n$ singularities in the Borel plane at
\ben
z= \frac{2+2n}{\beta_{1}}
\een

We can now use the stationary phase approximation in order to estimate $G_{\alpha}(g^{2})$, the result is
\ben
G_{\alpha}(g^{2}) = \frac{\Lambda^{4-2\alpha}}{(2-\alpha)} \frac{1}{g^{4}} \frac{1}{a} e^{- \frac{1}{g^{2}a}}
\een

Until now we have been working in four dimensions. The generalization to two dimensions
is quite simple. We get
\ben
G_{\alpha}(g^{2}) = \frac{\Lambda^{2-2\alpha}}{(1-\alpha)} \frac{1}{g^{4}} \frac{1}{a} e^{- \frac{1}{g^{2}a}}
\een
where now $\alpha < 1$. In next section we will use these expression in the analysis of
two dimensional non linear sigma models.

Finally we will extend our computations to the case in which we have $N=1$ supersymmetry
\footnote{In order to do that we can start replacing $P_{R}(k^{2})$
in (\ref{siete2}) by a superpropagator.}. The result 
in four dimensions is
\ben
G_{\hat \alpha}(g^{2}) = \frac{\Lambda^{4-2\hat \alpha}}{(2-\hat \alpha)} 
\frac{1}{g^{4}} \frac{1}{a} e^{- \frac{1}{g^{2}a}}
\een
where as before $a= \frac{\beta_{1}}{d}$, $d= 4- 2\hat \alpha$ and
\ben
\hat \alpha = \alpha + \frac{1}{2}
\een
with $\alpha < 2$. For $\alpha = 0$ this corresponds to $d=3$. For
$N=1$ super Yang Mills with 
\ben
\beta_{1} = \frac{3N}{8\pi^{2}}
\een
i.e $a=N$ we get
\ben\label{diez2}
G_{\hat \alpha = \frac{1}{2}}(g^{2}) = \Lambda^3 
\frac{1}{g^{4}} \frac{8\pi^{2}}{N} e^{- \frac{8 \pi^{2}}{g^{2}N}}
\een
that we can interpret as $<\lambda \lambda>$ gaugino condensate.

In the SUSY case the singularities in the Borel plane for $\hat \alpha = \frac{3}{2}-n$ are at
\ben
z= \frac{1+ 2n}{\beta_{1}}
\een

\subsubsection{ Fractional Instanton versus IR Renormalon}
Before reaching the main conclusion of this section concerning the connection
of IR renormalons and fractional instantons we must come back to our
computation of $G_{\alpha}(g^{2})$ and to fix the factors of $N$. In the case
of $N=1$ super Yang Mills with all the fields in the adjoint representation
it is natural to associate with $G_{\alpha}(g^{2})$ a factor $N^{2}$. Thus 
we must modify (\ref{diez2}) to
\ben
G_{\hat \alpha = \frac{1}{2}}(g^{2}) = \Lambda^3 
N \frac{1}{g^{4}} \frac{16\pi^{2}}{3} e^{- \frac{8 \pi^{2}}{g^{2}N}}
\een
Now we can compare this result with the fractional instanton result (\ref{tres2}) we formally got
by extending the index theorem to the case of fractional topological number. We observe that up to 
numerical factors we have not fixed in the formal instanton computation both nicely coincide.
In this way we can conclude the following result:

{\em For $N=1$ super Yang Mills the contribution of the first IR renormalon corresponding
to the Borel singularity at $z= \frac{3}{3N}= \frac{1}{N}$ is, once we estimate the Borel transform
in the stationary phase approximation, equivalent to a fractional instanton of topological number
$k= z= \frac{1}{N}$.}

Notice that in the $N=1$ case the IR renormalon
contribution produces the right powers of $g$, namely $\frac{1}{g^{4}}$ consistent
with the dimension of the moduli for topological charge $\frac{1}{N}$ \footnote{ Recall that 
the dimension of the moduli is $4Nk$ for $k$ the topological number}.

\subsubsection{Comments on Borel Summability}
In all the previous examples the IR renormalon singularity at
$z=\frac{1}{a}$ prevents ,a priori, to integrate in the Borel-Laplace transform from $z=0$ to $z=\infty$.
In fact the region where $F(z)$ is convergent is $z< \frac{1}{a}$ which is certainly not enough
to prove Borel summability. In our discussion above we have decided to estimate the Borel-Laplace
transform using a phase stationary approximation. This means that we have saturated
the integral over $z$ by the saddle point contribution. Recall from
(\ref{seis2}) that the saddle point is given by
\ben
z= \frac{1}{a} - g^{2}
\een
interpreted as a function $z(g^{2})$. Thus the contribution of the IR renormalon $z= \frac{1}{a}$ can be
obtained by taking the limit $g \rightarrow 0$ in the phase stationary result. It is instructive
to see that in this limit the leading contribution comes from the exponential factor
$e^{- \frac{8 \pi^{2}}{g^{2}N}}$ (in the case of $N=1$ SUSY Yang Mills ) that we can interpret as the
fractional instanton. Moreover this exponential
fractional instanton factor smooths the divergence.
At this point it is natural to ask why not to do the same for pure Yang Mills. Of course in pure
Yang Mills we have IR renormalons as well. The first one that can be associated with a gauge invariant
operator is at $z= \frac{4}{\beta_{1}}$ corresponding to $d=4$ and with
$\beta_{1} = \frac{11 N}{3}$. The exponential
factor in this case will becomes
\ben
e^{-\frac{8 \pi^{2} 12}{g^{2} 11 N}}
\een
which unfortunately can not be interpreted as a fractional instanton \footnote{ Formally it could be
interpreted as a fractional instanton of topological number $\frac{12}{11 N}$. Using index theorem this
will produce a dimension of moduli $4 N \frac{12}{11 N} = \frac{46}{11}$ which is certainly non sense.}.
All this seems to indicate that something dynamically very special 
take place when { \em the IR renormalon contribution
can be interpreted , using a formal extension of index theorems, as a fractional instanton}. It is
this coincidence what seems to be at the core of the magic of supersymmetry.

\subsubsection{Remark: On the $\eta'$ mass}
In reference \cite{wittentheta} it was assumed that for pure Yang Mills without fermions physics
depends on the $\theta$ vacuum parameter to order $\frac{1}{N}$ and not to order $e^{-N}$ as 
it is suggested by pure instanton effects. A direct consequencce of this assumption,
that was based on the solution in planar limit of two dimensional non linear sigma models \cite{Dadda}
\cite{wittemcp},
is a mass formula ,in the large $N$ limit, for the ninth light pseudoscalar, 
the $\eta'$ \cite{wittentheta}\cite{veneziano}. In fact in
the  presence of massless fermions physics should be independent of $\theta$ therefore in order
to cancell the $\theta$ dependence of pure Yang Mills we need a pseudoscaler of mass
$O(\frac{1}{N})$. 

The crucial quantity is the topological susceptibility that directly measures the $\theta$ dependence of
the vacuum energy of pure Yang Mills
\ben
U(k) = \int d^{4}x e^{ikx} <T(FF^{*}(x) FF^{*}(0)>
\een
in the limit $k=0$. This quantity is intimately related with IR renormalons. In fact we
have
\ben
<T(FF^{*}(x) FF^{*}(x)> = \int_{0}^{\Lambda} d^{4}k U(k) = 
\sum_{n}\int d^{4}k k^{4} (ln(\frac{\Lambda^{2}}{k^{2}}))^{n} C^{n} 
\een
with $C$ fixed by the beta function as usual. This is essentially equivalent to what we have
denoted $G_{\alpha}(g^{2})$ for $\alpha = -2$. The two dimensional version is
\ben
<T(FF^{*}(x) FF^{*}(x)> = \int_{0}^{\Lambda} d^{2}k U(k) = 
\sum_{n}\int d^{2}k k^{2} (ln(\frac{\Lambda^{2}}{k^{2}}))^{n} C^{n} 
\een
In both cases we can estimate this quantity using stationary phase approximation. The result
in four dimensions for Yang Mills is
\ben
<T(FF^{*}(x) FF^{*}(x)> \sim \Lambda^{8}\frac{1}{g^{4}}\frac{8 \pi^{2} 24}{11N} 
e^{-\frac{8 \pi^{2} 24}{11N g^{2}}}
\een
This of course is not exactly what we need in order to get the mass difference for the $\eta'$, that is given
in terms of $U(k=0)$. On the other hand using low energy theorems the following mass formula was derived in
\cite{shifman}
\ben
m_{\eta'}^{2} f_{\eta'} \sim <F(x)F(x)> 
\een
and we can try to estimate $<F(x)F(x)>$ as $G_{\alpha = 0}(g^{2})$, which is a good indication on the potential
contributions of IR renormalons to the $\eta'$ mass.

\section{VY Effective Lagrangians and Two Dimensional Non Linear Sigma Models}
\subsection{Non Supersymmetric Case: The $O(N)$ Model}
The large-N limit of the two dimensional $O(N)$ sigma model provides a simple
example of the IR renormalon. The model has $N$ vector fields $\sigma^a$
with a lagrangian
\ben
L = {1\over 2g^2} \sum_{a=1}^N (\partial \sigma^a)(\partial \sigma^a)
\label{eq:cone}
\een
with the constraint
\ben
\sum_a \sigma^a \sigma^a = 1
\label{eq:ctwo}
\een
This model can be exactly solved in the large-N limit by introducing
a lagrange multiplier field $\lambda (x)$ and rewriting the lagrangian
as
\ben
L = {1\over 2g^2} \sum_{a=1}^N (\partial \sigma^a)(\partial \sigma^a)
+ {N\lambda \over 2}(\sum \sigma^a\sigma^a - 1)
\label{eq:cthree}
\een
Integrating out the $\sigma$ fields now leads to an effective lagrangian
for the field $\lambda$
\ben
L = {N \over 2} {\rm Tr} \log (-{\partial^2 \over g^2} + N \lambda)
- {N \lambda \over 2}
\label{eq:cfour}
\een
At large-N, the functional integral over $\lambda$ is saturated by a
translationally invariant saddle point $\lambda_0$ which satisfies the
gap equation
\ben
\int {d^2p \over (2\pi)^2}{1 \over p^2 + g^2N \lambda_0} = {1\over g^2 N}
\label{eq:cfive}
\een
whose solution is
\ben
m^2 \equiv g^2N \Lambda_0 = \Lambda^2~{\rm exp}[-{4\pi \over g^2N}]
\label{eq:csix}
\een
where $\Lambda$ denotes an ultraviolet cutoff.
Expanding around this saddle point shows that $m^2$ is in fact the
dynamically generated mass of the field $\sigma^a$. This also leads 
to the exact beta function at $N = \infty$
\ben
\beta (g^2N) = \Lambda {d \over d\Lambda}g^2N = -2 {1\over 4\pi} (g^2N)^2
\label{eq:cseven}
\een
The running coupling constant is then given by
\ben
g^2 N(p^2) = {g^2 N \over 1 + {g^2 N \over 4\pi} \log ({p^2 \over \Lambda^2})}
\label{eq:ceight}
\een
Consider now the quantity
\ben
G_{\mu\nu}^{ab}(g^2N) = \int d(p^2) < \partial_\mu \sigma^a (p)
\partial_\nu \sigma (-p)>
\label{eq:ceighta}
\een
This is given by
\ben
G_{\mu\nu}^{ab}(g^2N) = -g^2 \delta^{ab}\int d(p^2){p_\mu p_\nu
\over p^2 + m^2}
\label{eq:cnine}
\een
Using the expression for the mass gap this evaluates to
\ben
G_{\mu\nu}^{ab}(g^2N) = -{g^2 \over 2}\delta^{ab}\delta_{\mu\nu}
\Lambda^2[1- {4\pi \over g^2N}e^{-{4\pi \over g^2N}}]
\een
Note that the first term is quadratically divergent while the second term,
which comes because of the presence of a mass gap is logarithmically
divergent. The latter can be seen by noting that the second term is
\ben
m^2 \log ({\Lambda^2 \over m^2})
\een
Subtracting the quadratic divergence one gets
\ben\label{GG}
(G^R)_{\mu\nu}^{ab} = {g^2 \over 2}\delta^{ab}\delta_{\mu\nu}
\Lambda^2~{4\pi \over g^2N}e^{-{4\pi \over g^2N}}
\een
 
The perturbation expansion of the model is performed by solving the
constraint explicitly
\ben
\sigma^N = {\sqrt{ 1 - \sum_i n^i n^i}}
\een
where we have renamed $\sigma^i = n^i$ for $i = 1 \cdots N-1$. The
action then has an infinite number of interaction terms
\ben
L = {1\over 2g^2}[\partial n^i \partial n^i + {n^in^j\partial n^i \partial
n^j \over (1 - n^in^i)}]
\een
and peforming the large-N expansion involves summation over an infinite
set of bubble diagrams. It is clear, however, that the final result
for the propagator is
\ben
<n^i (p) n^j (-p)> = \delta^{ij} {g^2 (p) \over p^2}
\label{eq:cten}
\een
where $g^2(p)$ is the running coupling constant defined in (\ref{eq:ceight}).

Let us now compute the quantity $G^{ij}_{\mu\nu}$ as defined in
(\ref{eq:ceighta}) by using (\ref{eq:cten}). The result is easily seen
to be 
\ben
G^{ij}_{\mu\nu} = {g^2 \over 2}\delta^{ij}\delta_{\mu\nu}
\Lambda^2~{4\pi \over g^2N}\int dy~{e^{-y}
\over {{4\pi \over g^2N} - y}}
\label{eq:celeven}
\een
It is now clear that the Borel transform $F_G(z)$ of this Green's function
has a pole at
\ben
z = \frac{4\pi}{N}
\een
This pole
is the infrared renormalon.
The integral in (\ref{eq:celeven}) is of course ill defined. If we
{\em define} this integral as being dominated by the pole we get a
result which is in agreement with (\ref{GG}). Since the 
latter is entirely due to mass generation, it is clear that the
dynamical mass may be thought of being produced by the IR renormalon.

We can now easily compare with the discussion on IR renormalons in previous section. In fact
we have
\ben
G^{ij}_{\mu\nu} = \delta^{ij}\delta_{\mu\nu}\int_{0}^{\Lambda}dk^{2} \frac{g^{2}}{1 + \frac{g^{2}N}{4\pi} 
log(\frac{k^{2}}{\Lambda^{2}})}
\een
which can be written as
\ben
\Lambda^{2} \int_{0}^{\infty} dz F(z) e^{- \frac{z}{g^{2}}}
\een
with
\ben
F(z) = \sum a^{n} z^{n} = \frac{1}{1-az} = \frac{1}{1- \frac{Nz}{4\pi}}
\een
where $a=\frac{\beta_{1}}{d} = \frac{N}{2\pi. 2}$. Since we are interested in 
the Green function after substracting the term of order $g^{2} \Lambda^{2}$ we should consider
\ben
\Lambda^{2} g^{2}\int_{0}^{\infty} dz F(z) e^{- \frac{z}{g^{2}}}
\een
Evaluating this integral using the stationary phase approximation we get a result in {\em perfect
agreement} with (\ref{GG}). This concludes the proof on the connection of IR renormalons and
mass generation in the $O(N)$ non linear sigma model.

\subsubsection{VY Effective Lagrangian}

Let us denote $\lambda$ the Lagrange multiplier. After gaussian integration
we get
\ben
L_{eff}(\lambda)dx^{2} = (- \frac{1}{g^{2}}\lambda + N 
log det||- \partial^{2} + \lambda||)dx^{2}
\een
Denoting
\ben
F(\lambda) = log det||- \partial^{2} + \lambda||
\een
we get
\ben
\frac{\partial F}{\partial \lambda} = 
\frac{1}{4\pi} log(\frac{\Lambda^{2}}{\lambda})
\een
which means
\ben
F(\lambda) = \frac{1}{4\pi} \lambda(log (\frac{\Lambda^{2}}{\lambda}) +1)
\een
thus
\ben
L_{eff}(\lambda)=- \frac{1}{g^{2}}\lambda 
+ \frac{N}{4\pi} \lambda(log (\frac{\Lambda^{2}}{\lambda}) +1)
\een
Defining
\ben
\Lambda^{2} = m^{2} e^{\frac{4\pi}{Ng^{2}}}
\een
for $m$ the mass scale and $\Lambda$ the cutoff, we get
\ben\label{uno}
L_{eff}(\lambda)= \frac{N}{4\pi}\lambda(log(\frac{m^{2}}{\lambda})+1)
\een
that is the VY effective lagrangian \cite{VY} for this model.

It is interesting to notice that we have obtained a VY effective lagrangian in a model
where instanton effects are manifestly absent.

\subsubsection{VY Effective Lagrangian, Renormalization Group and Legendre Transform}
For simplicity in notation let us define
\ben
x= \frac{1}{g^{2}}
\een
In terms of this variable let us define the function
\ben\label{zero}
f(x) = \frac{N}{4\pi} m^{2} e^{\frac{x}{N}}
\een
Let us now denote $\lambda$ the conjugated variable to $x$ and let us perform the Legendre
transform of $f(x)$, namely
\ben
Lf(\lambda) = \frac{N}{4\pi}\lambda ( log(\frac{m^{2}}{\lambda}) + 1)
\een
which is in fact the VY effective lagrangian (\ref{uno}). 
Recall that the Legendre transform $Lf(\lambda)$ of a function $f(x)$ is defined as
\ben
Lf(\lambda) = \lambda x(\lambda) - f( x(\lambda))
\een
for $x(\lambda)$ the solution to the Legendre equation
\ben\label{dos}
f'(x) = \lambda
\een
What we observe is the following nice fact. The function $f$ such that $Lf$ is the VY effective action
is just { \em the one such that the solution $x(\lambda)$ to equation 
(\ref{dos}) is the renormalization group
running coupling with $\lambda$ playing the role of the RG parameter.} In fact from (\ref{zero}) we get
\ben
x(\lambda) = \frac{1}{g^{2}}( \lambda) = \frac{N}{4\pi}log(\frac{\lambda}{m^{2}})
\een
recall that $\lambda$ , the Lagrange multiplier is of dimension $2$.

\subsection{The supersymmetric Case: $N=2$ Models}
The supersymmetric $CP^{N-1}$ model has been a typical toy model of four dimensional asymptotically
free field theories as QCD and $N=1$ super Yang Mills. Let us first briefly recall some well known facts
about this model. The $CP^{N-1}$ model is asymptotically free. The $n$ particles and the fermionic
superpartners $\psi$ associted with the elementary fields get dynamically a mass. Chiral condensates
$<\psi \bar \psi>$ get a vev with $N$ different vacua in agreement with the value
of $Tr(-1)^{F}$. There are solitons interpolating between the different vacua that are 
the massive $n$ and $\psi$ particles. The quantum chiral ring has the structure $x^{N}=1$. The model
possesses a $N=2$ supersymmetry and the solitonic spectrum admits a $N=2$ Landau Ginzburg description. 
The chiral anomaly is 
$\partial j_{\mu}^{5} = 2Ng^{2}\epsilon_{\mu, \nu} \partial_{\mu} \bar n \partial_{nu} n$ which means
that as well as it is the case for $N=1$ super Yang Mills we can expect a $\eta'$ of mass of order one.

In order to fix notation let us briefly review some generalities on two dimensional $N=2$ models.
We will denote $\Phi_{i}$ the chiral superfields and $V$
the vector superfield. Under $U(1)$ gauge transformations
\ben
V \rightarrow V + i( \Lambda - \bar \Lambda)
\een
with $\Lambda$ a chiral superfield, and
\ben
\Phi_{i} \rightarrow e^{-i Q_{i} \Lambda} \Phi_{i}
\een
for $Q_{i}$ the $U(1)$ charges. We will reduce ourselves to abelian gauge group.
It is convenient to introduce the superfield $\Sigma$ as
\ben
\Sigma = \frac{1}{\sqrt{2}}\bar D_{+} D_{-} V
\een
We will denote $\sigma$ the scalar component of this superfield.
The lagrangian is given by
\ben
L = \sum_{i} \Phi_{i}e^{Q_{i} V} \Phi_{i} + W({\Phi}) + \frac{1}{4 e^{2}} \bar \Sigma \Sigma
- r V
\een
where $W$ is the standard superpotential, $e$ is the gauge coupling constant and
$-r V$ is the Fayet Iliopoulos term with $r$ a free parameter. We can also add
to this lagrangian the usual  $\theta$ term. The parameters $r$ and $\theta$ combine
into a complex variable $t$ defined as
\ben
t= ir + \frac{\theta}{2 \pi}
\een
Using $t$ the FI term and the $\theta$ term can be written as
\ben\label{SW}
i\frac{1}{2 \sqrt{2}} t \Sigma + c.c
\een
From now on we will consider for simplicity models with $W=0$. Using the
equations of motion we get
\ben
D = - e^{2}(\sum_{i}Q_{i} |\phi_{i}|^{2} - r)
\een
where $D$ as usual is the last component of the vector superfield $V$ and$\phi_{i}$
the scalar components of the chiral superfields $\Phi_{i}$.
The classical potential is
\ben
U = \frac{1}{2 e^{2}} D^{2} + 2|\sigma|^{2} \sum_{i} Q_{i}^{2}|\phi_{i}|^{2}
\een
Let us now consider the classical supersymmetric vacua defined by $U=0$. 
If the FI coupling $r$ is non vanishing the solution is $\sigma = 0$ and
\ben\label{seven4}
\sum_{i}Q_{i} |\phi_{i}|^{2} = r
\een
Around this vacua the $\sigma$ field becomes massive and the low energy
lagrangian is simply the sigma model with target space defined by (\ref{seven4}). If we chose
$Q_{i}=1$ and $\sum_{i}Q_{i} = N$ we get from (\ref{seven4}) the standard
supersymmetric $CP^{N-1}$ model.
On the other hand if we consider $r=0$ we get a diffeterent type of solution to $U=0$, namely
$\phi_{i}=0$ and $\sigma$ different from zero. Around this classical vacua
the $\sigma$ field is massless but the chiral superfields $\Phi_{i}$ become massive
with a mass of the order of $|\sigma|$. The corresponding low energy physics
would be obtained after integrating these massive fields \cite{wittenN}, \cite{silverstein}.

Let us briefly describe the integration of the massive fields. For the $CP^{N-1}$ model
with $N$ chiral superfields and charges $Q_{i} = 1$
we get
\ben
\int d\phi_{i} exp - \int (D\phi_{i} D\bar \phi_{i} + D(|\phi_{i}|^{2}) + |\sigma|^{2}|\phi_{i}|^{2})
= exp - L_{eff}(D,\sigma)
\een
Performing the integration we get
\ben
L_{eff}(D,\sigma) = \frac{1}{4\pi}N((D+ |\sigma|^{2}) log( \frac{\Lambda^{2}}{(D+ |\sigma|^{2})}) + D 
+ |\sigma|^{2}
\een
After adding the tree level term $-rD$, the equation of motion for the auxiliary field $D$ is
\ben\label{seis}
\frac{1}{4\pi} N log(\frac{\Lambda^{2}}{D + |\sigma|^{2}}) -r = 0
\een
Now we can ask ourselves what is the effective action $L(\Sigma)$ that reproduces this equation of motion for the
$D$ field. It is easy to observe that for large $|\sigma|^{2}$ the F-term is given by
\ben
L_{F}(\Sigma) = d\theta^{2} (\frac{N}{2\pi}N\Sigma(log(\frac{\Lambda}{\Sigma}) + 1) +
\frac{it}{2 \sqrt{2}} \Sigma)
\een
with $t$ at the scale $\Lambda$.
This is the VY effective lagrangian as it should be expected.

For future use we will consider also the case of the $N=2$ model associated to the conifold. In this case
we have $4$ fields with charges $+1,+1,-1,-1$. The effective action
$L_{eff}(D, \sigma)$ in this case is
\ben\label{ocho}
L_{eff}(D, \sigma) = \frac{1}{4\pi} ( 2 (D + |\sigma|^{2}) log( \frac{\Lambda^{2}}{(D+ |\sigma|^{2})})
+ 2( |\sigma|^{2} - D) log( \frac{\Lambda^{2}}{|\sigma|^{2} - D}) + 4 |\sigma|^{2})
\een
After adding the tree level terms $-rD + D^{2}$ the equation of motion for $D$ is
\ben
D + \frac{1}{4\pi} (2 log(\frac{(|\sigma|^{2} - D)}{(D+ |\sigma|^{2})}) -r = 0
\een
Expanding in $|\sigma|^{2}$ we get
\ben\label{nueve}
D( 1- \frac{1}{\pi |\sigma|^{2}}) = r
\een
The F-term of the corresponding effective lagrangian $L(\Sigma)$ is just $it \Sigma$. This reproduces the
equation $D=r$. In order to reproduce the correction $\frac{1}{|\sigma|^{2}}$ in (\ref{nueve}) we need to add
to the effective lagrangian a D-term of the type \cite{Davis}, \cite{silverstein}
\ben\label{diez}
log \Sigma log \bar \Sigma d\theta^{2} d\bar \theta^{2}
\een
The same is true for the equation of motion (\ref{seis}) for $CP^{N-1}$ model. Corrections of  
$O(\frac{D}{|\sigma|^{2}})$ to $log(\frac{\Lambda^{2}}{|\sigma|^{2}})$ generate an extra D-term
as the one given in (\ref{diez}). These comments will become relevant in next subsection.

Notice from (\ref{ocho}) that although the $N=2$ model associated with the conifold is conformal invariant
with vanishing beta function the effective lagrangian for $\sigma$ if $D=0$ is again a VY effective lagrangian.

As we did for the $O(N)$ model we can reproduce these results from the point of view of IR renormalons.
In fact in this case we should just consider in two dimensions $G_{\hat \alpha = \frac{1}{2}} (g^{2})$ 
with $\beta_{1}= \frac{N}{4\pi}$. This
IR renormalon contribution reproduces $<\Sigma>$. As before their Legendre transform
reproduces the VY effective action.

\subsubsection{Topological Susceptibility}
As a final comment let us say few words on the topological susceptibility for the $CP^{N-1}$ model.
At finite $N$ this model has instantons which leads to a $\theta$
dependence of the vacuum energy. In the $\theta = 0$ the 
signature of this fact is the behavior
of the  
two point function of the topological charge density $T(x)$
\ben
T(x) = {1\over 2\pi}\epsilon_{\mu\nu}\partial_\mu A_\nu
\een
at zero momentum. It is clear that to any finite order in
perturbation theory the quantity
\ben
{\rm Lim}_{p \rightarrow 0} <T(p) T(-p)>
\een
vanishes since each $T$ is a total derivative. However in the presence
of instantons, a standard calculation based on dilute gas approximations
leads to a $\theta$ dependence of the vacuum energy of the model, and
hence to a nonvanishing topological susceptibility. The result
goes as ${\rm exp}~[-{1\over g^2}]$

In the large-N limit, $N \rightarrow \infty, ~g \rightarrow 0$ with
$g^2N$ fixed instantons give a vanishing contribution and one might think
that in this limit the vacuum energy becomes independent of $\theta$.
However, as has been shown in the 80's this conclusion is wrong. In
a way entirely similar to the $O(N)$ nonlinear sigma model the 
lagrange multiplier acquires a nonzero
vacuum expectation value. Furthermore, 
expanding around the saddle point one finds that the auxiliary
gauge fields acquire a kinetic term
\ben
{g^2 N \over 48 \pi M^2} (\partial_\mu A_\nu - \partial_\nu A_\mu)^2
\een
A straightforward calculation of the two point function of $T(p)$ then
leads to the result
\ben
<T(p) T(-p)> = {3 M^2 \over \pi g^2 N}
\een
The crucial ingredient in this is the emergence of a kinetic term for
the gauge field. The propagator then cancels the explicit power of
momentum which comes from the definition of $T(p)$.

In the $N=2$ formalism, the dynamically generated kinetic term for the
auxiliary gauge field is contained in the D-term (\ref{diez}) discussed above. Recall that
we find this term when we consider effects of $O(\frac{D}{|\sigma|^{2}})$ which is the analog 
of expanding around the saddle point. From the IR renormalon point
of view we were able to derive the F-term corresponding to VY effective lagrangian but
not the D-term.

\section{VY and large $N$ Duality: Some Comments}
\subsection{The stringy meaning of VY effective Lagrangian}
It is a well known result of the large $N$ limit that the
perturbative expansion of quantum field theory
amplitudes for gauge theories is organized 
as a stringy genus expansion where
the gauge coupling $g^{2}$ is playing the role of the
string coupling. Generically we expect for the perturbative
expansion the following structure
\ben\label{cero}
\sum_{l} (g^{2})^{2l -2} F_{g}(t)
\een
where $t=g^{2}N$ is the t'Hooft coupling and $l$ represent the genus. 
On the other hand we know
that renormalon effects survive at large $N$ and we expect that they are
important contributions in the planar $l=0$ limit. For $N=1$
supersymmetric Yang Mills we have computed the renormalon contribution in 
section two ( see equation (20)) and we have derived, using
Borel transformation and stationary phase approximation, the following 
result for $d=3$
\ben
G_{\hat \alpha = \frac{1}{2}}(g^{2}) = \Lambda^{3} N^{2} \frac{1}{g^{4}} \frac{8\pi^{2}}{N}
e^{-\frac{8\pi^{2}}{g^{2}N}}
\een
If now we try to read this contribution as the genus zero contribution
in the sense of (\ref{cero}) we observe that the counting
of powers of $g$ is just the right one and we will get
\ben
\frac{1}{g^{2}} F_{0}(t)
\een
with
\ben
F_{0}(t) = \Lambda^{3} N 8\pi^{2}
e^{-\frac{8\pi^{2}}{t}}
\een
This is quite nice since it is exactly the Legendre transform of the
VY effective lagrangian, where we take $S$ as the conjugated variable.

In summary we observe two things. The first one
is that the VY effective F term lagrangian is just the Legendre transform
of the renormalon contribution to the genus zero amplitude. The second one
is that the factor $\frac{1}{g^{4}}$ corresponding
to the genus zero contribution is precisely the one fixed by the
renormalon contribution and also the one you will expect, by naiv 
counting of zero modes ( see discussion in section two ) for a fractional
instanton configuration of topological number $\frac{1}{N}$. Recall that
in this case and independently of the rank of the gauge group we get just
the four translation zero modes. This seems to indicate that the so called
fractional instantons have a very natural interpretation as a zero genus 
contribution.

There is a formal connection of VY with the $N=2$ model associated
with the conifold \cite{vafaooguri} as well as with the $c=1$ string. If we consider
the Laplace transform
\ben\label{tres7}
F(S) = \int_{0}^{\infty} \frac{1}{\sigma^{2}} e^{-S \sigma}
\een
this is divergent for ${\sigma \sim 0}$. This is exactly the same
problem we find in the $c=1$ case for the contribution
of surfaces of small area \cite{Gross} . The standard way to cure this
divergence is by diferenciating with respect to $S$. After performing 
two derivatives we get
\ben
\frac{\partial^{2} F}{\partial^{2} S} = \frac{1}{S}
\een
and therefore $F(S)= S log S$
i.e the structure of VY effective lagrangian. The integrand in (\ref{tres7})
can be on the other hand naturally associated with the $N=2$ conifold
model ( see section 4.2) if we identify $S$ with the FI coupling.

\subsubsection{Toroidal Compactification}
Using holomorphicity we can read the F-terms of $N=1$ super Yang Mills
from the F- terms of the $N=2$ theory we obtain by compactification on $T^{2}$ \cite{vafa2}.
For the case of $U(N)$ the two dimensional model
is the $N=2$ non linear sigma model with quantum chiral ring isomorphic
to the chiral ring of the original four dimensional field theory, namely
the $N=2$ $CP^{N-1}$ model we have briefly described in section 3. The Kahler
class is identified with the Yang Mills coupling constant, and
$\Sigma$ with the ``dimensional reduction'' of the glueball field $S$.

The isomorphism of the chiral rings between the two models is quite clear at
the level of the renormalon contributions. In four dimensions we get
\ben
\int \frac{d^{4}k}{k^{2 \alpha_{2}}} (log k^{2})^{n}C_{4}^{n}
\een
with $C_{4}=\frac{\beta_{1}}{2}$ for $\beta_{1}= 3N$
In two dimensions
\ben
\int \frac{d^{2}k}{k^{2 \alpha_{2}}} (log k^{2})^{n}C_{2}^{n} 
\een
with $C_{2}= N$. Clearly both are the same if in four dimensions we consider
an operator of dimension $3$ and in two dimensions an oparator of 
dimension $1$. More precisely the condition for this dimensional 
reduction to work at the renormalon level
is
\ben
\frac{d}{\beta_{1}^{D=2}} = \frac{d+2}{\beta_{1}^{D=4}}
\een
where $\beta_{1}^{D}$ is the beta function for the D-dimensional
theory. For $d=2$ i.e non supersymmetric case this equation have in general
no solution.

\subsection{Renormalons and Matrix Models}
Let us consider a formal perturbative expansion
\ben
\sum_{n} g^{2n} C^{n} n!
\een
where $C$ is given in terms of the beta function. This serie diverges
as soon as the $N+1$ term becomes larger than the $N$th. This take place when
\ben\label{uno3}
g^{2} C n = 1
\een
We can cutoff the perturbative series at the order determined by (\ref{uno3}). The 
error introduced by this cutoff can be estimated as the value of $g^{2n} C^{n} n!$ for
\ben\label{uno4}
n= \frac{1}{Cg^{2}}
\een
We will identify this estimate as the {\em non perturbative contribution}.
Using the asymptotic Stirling's formula
\ben
log\Gamma(n) = (n+\frac{1}{2})ln n -n + \frac{1}{2} ln 2\pi + \sum_{l} \frac{B_{2l}}{2l (2l -1) n^{2l-1}}
\een
we get in first approximation
\ben
g^{2n} C^{n} n! = g^{2n} C^{n} n^{n} e^{-n} = e^{- \frac{1}{g^{2}C}}
\een
where we have used (\ref{uno4}). 
For $C= \frac{N}{8 \pi^{2}}$ we get the typical
fractional instanton exponential.
Notice that
if we use the whole expansion including the Bernoulli numbers we will get
\ben
e^{- \frac{1}{g^{2}C} + \sum_{l} \frac{B_{2l} (g^{2} C)^{2l-1}}{2l (2l -1)}}
\een
that
can be probably interpreted as a hidden gravitational correction to the fractional instanton
contribution.
Let us define $F_{np}$ as
\ben
F_{np}(n) = ln( g^{2n} C^{n} n!) \sim ln ( (ng^{2}C)^{n} . e^{-n})
\een
estimated at $n>\frac{1}{Cg^{2}}$ using Stirling's formula. Now we look for
a sort of {\em``prepotential''} for $F_{np}$ i.e some functional $\Phi(n)$ such that
\ben
\frac{\partial \Phi(n)}{\partial n} = F_{np}(n)
\een
In the asymptotic regime the logarithm of Barnes $G$ function is a natural candidate
for the {\em ``prepotential''} $\Phi(n)$ \footnote{ This is very much related with Malmsten's formula.}. 
In fact
the asymptotic expansion of Barnes function is
\ben\label{cuatro4}
ln G(1+n) = n^{2}(\frac{ln n}{2} - \frac{3}{4}) + \frac{ln(2\pi)}{2}n - \frac{1}{12} ln n + \zeta'(-1) -
\sum \frac{B_{2k+2}}{4k (k+1) n^{2k}}
\een
Therefore
\ben
\frac{\partial \frac{1}{g^{4}C^{2}} ln G(1 + n g^{2}C)}{\partial n} \sim n( ln(ng^{2}C) -1) \sim F_{np}(n)
\een
which means that we can take as ``prepotential'' $\Phi(n)$ for the non perturbative contribution
$\frac{1}{g^{4}C^{2}} ln G(1 + n g^{2}C)$.

The logarithm of Barnes $G(1+\hat N)$ function is intimately related with the partition
function of $U(\hat N)$ Chern-Simons
\cite{vafaooguri} on $S^{3}$ and with
the gaussian Matrix model for $\hat N \times \hat N$ matrices \cite{vada dijk}, \cite{cachazo}. 
In both cases the contribution
of factorials comes from the volume of $U(\hat N)$ in the large $ \hat N$ limit. 
Moreover the {\em genus} expansion
of the logarithm of Barnes $G$ function coincides with the 
free energy of Penner model \cite{Vafa Distler}. 

What we have observed in the previous exercise
is that the natural connection with four dimensional supersymmetric gauge theories is
essentially  due to the fact that the most natural estimate of the non perturbative effects
comes from the contribution of the $n!$ ( i.e generic renormalon effects) in the
perturbative expansion. 
Notice that this can be done independently if the theory is or not supersymmetric. However supersymmetry
could be crucial in order to relate the $n!$ i.e the $\Gamma$ function of the perturbative expansion
with Barnes $G$ function that play the formal rol of a prepotential. A much more deep analysis of
this relation \cite{work} is necessary but all this seems to indicate that some topological field theories or
$c=1$ Matrix models are playing the role of bookkeepings of the $n!$ renormalon divergences of
asymptotically free massless field theories.
 
\subsubsection{Remark: Hidden Gravity?}
The logic of the previous section was based on estimating the error when we cutoff the 
perturbative expansions
even in the very weak coupling regime. In asymptotically free theories and when we consider
typical renormalon divergences as we did in previous section we find quite naturally what looks as
a gravitational contribution, namely
\ben
e^{- \frac{1}{g^{2}C}(1 - \sum_{l} \frac{B_{2l} (g^{2} C)^{2l}}{2l (2l -1)})}
\een
In the IR renormalon case with $C$ determined by the beta function we get ,in the $N=1$
supersymmetric case, a formal expansion in t'Hooft coupling $t$. The relevant {\em''gravitational''}
correction is given by
\ben\label{cuatro3}
\sum_{l} \frac{B_{2l} (t)^{2l-1}}{2l (2l -1)}
\een
As a formal infinite series this 
is highly divergent even for very small $t$ due to the grow of the
Bernoulli numbers as $2l!$. This is similar to the familiar situation in
Matrix $c=1$ models (\cite{grosss}). \footnote{A Borel resummation can be defined as
$\sum_{l} \frac{B_{2l} (t)^{2l-1}}{2l (2l -1)} = \int_{0}^{\infty}dz F(z) e^{-\frac{z}{t}}$
with
$F(z) = \frac{1}{z}(\frac{1}{2} -\frac{1}{z} + \frac{1}{e^{z} -1})$
which has singularities at $z= 2\pi i n$.}. Notice that in (\ref{cuatro4}) we really have 
$\sum^{m} \frac{B_{2k+2}}{4k (k+1) n^{2k}} + O(\frac{1}{n^{2m+2}})$.

We can also consider divergences of {\em instanton} type if we are not working in the planar limit.
In this case the constant $C$ is just a number of the order $8\pi^{2}$ and (\ref{cuatro3})
becomes a expansion in $g_{YM}^{2}$ i.e in string coupling constant similar
to the standard genus expansion.

In summary we observe that for asymptotically free theories the uncertainties due to 
renormalon ( and instanton) divergences in the perturbative expansion can be interpreted as 
associated with
a hidden gravitational sector in the sense of Matrix $c=1$ models. Many years ago
t' Hooft was suggesting that asymptotic freedom is probably not enough \cite{tHooftPHyRe}. He was 
refering to the uncertainties in the perturbative expansion even for arbitrarily small coupling. It
looks that the new ingredient we need to fix these uncertainties is intimately related with
the gravitational sector hidden in asymptotically free gauge theories.

\section{Acknowledgements}
I would like to thank Sumit Das for collaboration in the early stages of this work and
for many crucial comments and suggestions. I would like also to thank Pepe Barbon and Rafa Hernandez 
for valuable
discussions. This research was supported by Plan Nacional de Altas Energias, Grant FPA2003-02-877

\end{document}